\documentclass[journal]{IEEEtran}

\usepackage{cite}
\usepackage{graphicx}

\ifCLASSINFOpdf

\else

\fi

\usepackage{authblk}

\begin{document}

\title{Time response of a microring resonator to a rectangular pulse in different coupling regimes}

\author[1*]{Stefano Biasi}
\author[1]{Pierre Guillem\'e}
\author[1]{Andrea Volpini}
\author[1]{Giorgio Fontana}
\author[1]{Lorenzo Pavesi}
\affil[1]{\small{Nanoscience Laboratory, Departement of Physics, University of Trento, Via Sommarive 14, Povo - Trento, 38123, Italy}}
\affil[*]{\small{E-mail: stefano.biasi@unitn.it}}

%\author{Stefano~Biasi,
 %       Pierre~Guillem\'e, Andrea~Volpini, Giorgio~Fontana
  %      and~Lorenzo~Pavesi% <-this % stops a space
%\thanks{S. Biasi, P. Guillem\'e, A. Volpini, G. Fontana and L. Pavesi are with the Nanoscience Laboratory, Departement of Physics, University of Trento, Via Sommarive 14, Povo - Trento, 38123, Italy e-mail: (stefano.biasi@unitn.it).}% <-this % stops a space

%\thanks{}}

% The paper headers
%\markboth{Journal of \LaTeX\ Class Files,~Vol, No}%
%{Shell \MakeLowercase{\textit{et al.}}: Bare Demo of IEEEtran.cls for IEEE Journals}

% make the title area
\maketitle

% As a general rule, do not put math, special symbols or citations
% in the abstract or keywords.
\begin{abstract}
We discuss the analytical temporal response of a microring resonator excited through a bus waveguide by an optical rectangular pulse. Finite difference time domain (FDTD) simulations illustrate the analytical solution and help in understanding the meaning of the different coupling regimes.
In addition, we show that the temporal dynamics allows determing the coupling regime while the commonly used spectral characterization in the stationary regime does not.
We also take advantage of the simulation to highlight the phase shift between the input and the output signals in the different coupling regimes. Finally, measurements on a  $Si_3N_4$ microring resonator are performed and analyzed in the case of under-coupling regime
to illustrate how the time response study leads to the Q-factor determination.
\end{abstract}

% Note that keywords are not normally used for peerreview papers.

\IEEEpeerreviewmaketitle

\section{Introduction}

\IEEEPARstart{M}{icrorings} coupled to a bus waveguide are commonly used in integrated photonics \cite{heebner_optical_2008}. Their light storage capacity is given by means of the quality factor (Q). Several experimental techniques have been developed to estimate this coefficient ranging from the simple measurement of the frequency transmission response \cite{Ramiro-Manzano:12,BogaertsM} to more complex cavity-ring-down spectroscopy techniques \cite{Anderson:84,cavitybook,Barnes:08,Armani2003}. The first one is easy to implement, but, contrarily to the second one, it does not give detailed information on the coupling regime. In fact, the bus waveguide can be coupled to the microring in three characteristic regimes: under-coupled, over-coupled and critical-coupled. Among these, the critical coupling regime is characterized by the fact that no light is detected at the bus waveguide output when a continuous signal is input at the bus waveguide. It is often said that all the light is coupled into the microresonator \cite{Rauschenbeutel,Kippenberg}. Since the critical coupling is defined as the regime where the injected power exactly compensates the losses inside the resonator, and since these losses are relatively weak in microresonators, the fact that all the pump power enters into the resonator seems contradictory.  

This paper aims at clarifying this apparent contradiction by illustrating the propagation of a rectangular pulse of light in the system. This pulsed excitation scheme allows characterizing the coupling regime of the system just looking at the intensity of the time dependent electric field on a resonance of the microring. Moreover, we discuss how the response of the system is determined by the interference at the coupling region between the light coupled to the resonator and that propagating in the bus waveguide. The different coupling regimes show different interference conditions which determine both the phase delay of the output light as well as the temporal lineshape of the output pulse. Finally, we propose an analytical complex expression for the single Lorentzian transmission response which can be used to fit the experimental data and, therefore, to extract the characteristic parameters of the system, e.g. the Q-factor, the intrinsic and extrinsic coefficients.

\section{Theory and simulation} 
\label{Sec:Theory}
\subsection{Analytical model development} \label{sec:generalResolution} 
% ----------------------------------------------

\begin{figure}[!t] %%ht!
	\begin{center}
		\centering\includegraphics[width=7.5cm]{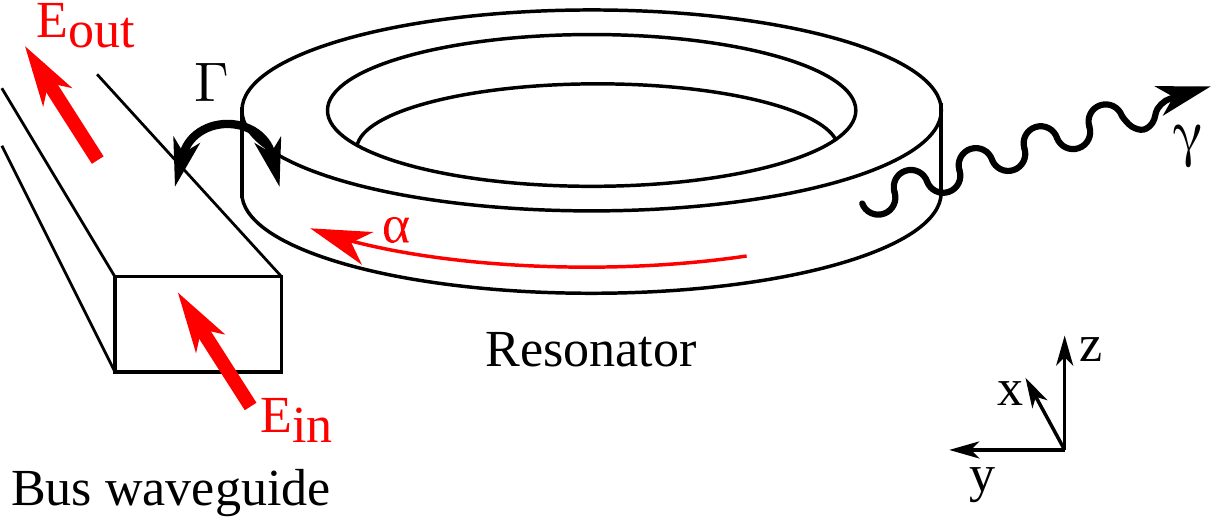}
		\caption{Sketch of a resonator coupled to a bus waveguide. $\Gamma$ ($\gamma$) extrinsic (intrinsic) damping rate (i.e. losses). $E_{in}$, $E_{out}$ and $\alpha$ are the electric field amplitudes respectively at the input of the bus waveguide, at its output and in the resonator.}
		\label{fig:system}
	\end{center}
\end{figure}

Let us consider a single mode resonator coupled to a bus-waveguide (see figure \ref{fig:system}). An electric field, with amplitude $E_{in}$, injected into the input port of the waveguide couples into the resonator and excites a mode whose electric field amplitude is $\alpha$. The Temporal Coupled Mode Theory (TCMT) \cite{Joann:1,Milburn:1,Laine:1,Shanhui:1,Yariv} and the usual properties of a time-reversal-invariant system lead to:

\begin{eqnarray}
	\frac{d\alpha}{dt} &=&(i\omega_{0}-\gamma-\Gamma)\,\alpha[t]+i \sqrt{2\Gamma} E_{in}[t] \label{eq:eqDiff}\\
	E_{out}[t] &=& E_{in}[t] + i \sqrt{2 \Gamma} \; \alpha[t] \label{eq:Eout},
\end{eqnarray}

where $\omega_0$ is the resonant angular frequency of the resonator, $E_{out}$ is the output electric field amplitude and $\Gamma$ ($\gamma$) is the extrinsic (intrinsic) damping rate which are positive real values. Let us note that $\gamma$ describes the losses in the resonator which are due to intrinsic factors (material absorption, scattering, bending, ...) while $\Gamma$ is related to the coupling with the bus waveguide (extrinsic factor).

Considering a monochromatic input field $E_{in}[t]=E_{in0}[\omega]\,e^{i\omega t}$ one can obtain the analytical expression of the field amplitude transmission $t_r = E_{out}/E_{in}$:

\begin{equation} \label{eq:tr}
t_r=1-\frac{2\Gamma}{i\,(\omega - \omega_{0})+\gamma+\Gamma} = 1 - \frac{2\Gamma}{i \Delta \omega + \gamma + \Gamma},
\label{eqa:TrasmissionResponse}
\end{equation}    
where we have defined the detuning $\Delta \omega=\omega-\omega_{0}$. The plot of the square modulus of this expression gives a Lorentzian shape, which is the typical \textit{frequency response} of a resonant system, as we will see in subsection \ref{sub:frequency}. 

On the other hand, to compute the \textit{time response}, we have to solve equation \ref{eq:eqDiff}. 
To keep the resolution general, we use the Green's function $G [t]$, which is defined as the solution of a differential equation when its forcing equals a Dirac's function pulse $\delta$ \cite{GreenBook}. The general solution is, then, given by the convolution of $G [t]$ with the exciting function.
Taking the Fourier transform of equation \ref{eq:eqDiff} doing the substitution $E_{in}[t] = \delta[t-t']$ one obtains:
\begin{equation}
i \omega \; G[\omega] - (i \omega_0 - \gamma - \Gamma) G[\omega] = i \frac{\sqrt{2 \Gamma}}{\sqrt{2\,\pi}} \; e^{i \omega t'}.
\end{equation}
In this way, the spectral Green's function of equation \ref{eq:eqDiff} reads as:
\begin{equation}
G[\omega] =\frac{1}{\sqrt{2\,\pi}} \frac{i \sqrt{2 \Gamma}}{i(\omega - \omega_0)+\gamma + \Gamma} \; e^{i \omega t'}.
\end{equation}
The inverse Fourier transform leads to the Green's function:
\begin{equation} \label{eq:G}
G[t-t'] = i \sqrt{2 \Gamma} \; e^{-(-i \omega_0 + \gamma + \Gamma)(t-t')} \; \Theta[t-t'],
\end{equation}
where $\Theta [t]$ is the Heaviside function. Finally, the general solution of equation \ref{eq:eqDiff} reads as:
\begin{equation}
\alpha[t] = \int_{-\infty}^{\infty} G[t-t'] \; E_{in}[t'] \; dt'.
\end{equation}
Therefore, the general response of the system is given by substituting the last expression in equation \ref{eq:Eout}:
\begin{equation} \label{eq:EoutSolution}
E_{out}[t] = E_{in}[t] + i \sqrt{2 \Gamma} \int_{-\infty}^{\infty} G[t-t'] \; E_{in}[t'] \; dt'.
\end{equation}

% ------------------------------------------------------------------------------------
\subsection{A time dependent rectangular pulse as input field}
\begin{figure}[!t] %%ht!
	\begin{center}
		\centering\includegraphics[width=8.1cm]{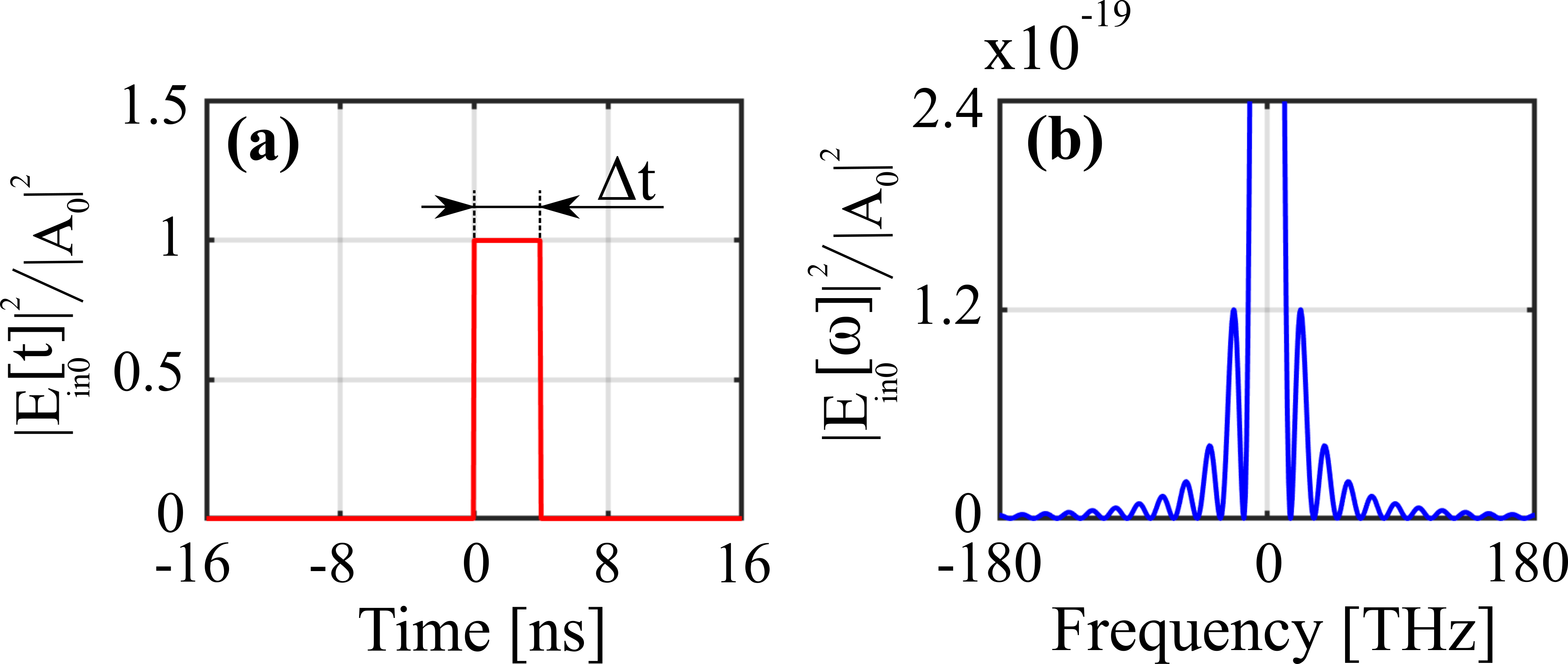}
		\caption{(a) Time dependent rectangular pulse envelope and its Fourier transform (b).}
		\label{fig:InputField}
	\end{center}
\end{figure}

Let us assume that the resonator is excited by a continuous wave (CW) pulse of light of duration $\Delta t$ and of angular frequency $\omega$ (see figure \ref{fig:InputField} (a)). Thus, the input field reads as:

\begin{equation} \label{eq:Ein}
E_{in}[t]=A_0\,\left( \Theta[t]-\Theta[t-\Delta t] \right) \; e^{i \omega t},
\label{eqa:StepFunct}
\end{equation}
where $A_0$ is the amplitude of the input field. As shown in figure \ref{fig:InputField}, the Fourier transform of the excitation envelope $E_{in0}[t]$ is a cardinal sine (Sinc) response:

\begin{equation}
\begin{array}{r@{\ }c@{\ }l}
E_{in0}[\omega]&=&\frac{1}{\sqrt{2\pi}} \int_{-\infty}^{\infty}E_{in0}[t]\,e^{-i\,\omega\,t} \\
& \vdots & \\
&=&\frac{i A_0 \left(e^{-i\,\omega\,\Delta t} - 1\right)}{\sqrt{2\pi}\omega }.\\
\end{array}
\label{eqa:SincFunct}
\end{equation}
As expected, from the last equation, one can see that by decreasing the duration of the pulse ($\Delta t$), the frequency bandwidth of the Sinc increases. 

Substituting the expression \ref{eq:Ein} in equation \ref{eq:EoutSolution} and using the Green's function \ref{eq:G} one obtains the response of the resonator:

\begin{equation} \label{eq:EoutGeneral}
\begin{array}{r@{\ }c@{\ }l}
E_{out}[t]& = &  A_0 e^{i \omega t} \left\{- \Theta[t]\left( \frac{-2 \Gamma e^{-t(i \Delta \omega + \Gamma+\gamma)}}{i\Delta \omega + \Gamma+\gamma} \right. \right. \\
& + & \left. \left. \frac{(\Gamma-\gamma)-i\Delta \omega}{i\Delta \omega + \Gamma+\gamma}\right) + \Theta[t-\Delta t] \left(\frac{(\Gamma-\gamma)}{i \Delta \omega +\Gamma+\gamma} \right. \right.\\
& - & \left. \left.  \frac{2 \Gamma e^{-(t-\Delta t)(i \Delta \omega + \Gamma+\gamma)}+i \Delta \omega}{i \Delta \omega +\Gamma+\gamma} \right) \right\}.\\
\end{array}
\end{equation}

This general expression will be discussed in section \ref{sec:freqDependence}. In the case of a resonant excitation ($\omega = \omega_0$), $E_{out}[t]$ reduces to:

\begin{equation}
\begin{array}{r@{\ }c@{\ }l}
E_{out}[t]& = &  A_0 e^{i \omega_0 t} \left\{- \Theta[t]\left( \frac{-2 \Gamma e^{-t(\Gamma+\gamma)}+(\Gamma-\gamma)}{\Gamma+\gamma}\right) \right.\\
& + & \left.  \Theta[t-\Delta t] \left(\frac{-2 \Gamma e^{-(t-\Delta t)(\Gamma+\gamma)}+(\Gamma-\gamma)}{\Gamma+\gamma}\right)\right\}. \\
\end{array}
\label{eqa:Eout}
\end{equation}
This simple equation describes the temporal dependence of the output pulse on the resonator parameters ($\omega_0$, $\Gamma$, $\gamma$).

\subsection{Gaussian pulse as input field}

In this work, we focus the attention on the case of a rectangular input field because, as far as we know, it has not been described in detail yet and, above all, it is a didactic example where the finite-difference time-domain (FDTD) illustrations can be readily understood. Nevertheless, to make the link with more developed works, in this section, we present how the equations would change  in the case of a Gaussian pulse. The dynamics of such an excitation within a cavity is well described in literature and is needed to understand important applications such as cavity dumping \cite{IntroLaserTech} and high-speed optical communication \cite{Jin2015}.

Let us consider the case where the system is resonantly excited through a Gaussian pulse of the form:

\begin{equation}
E_{in}[t]=\frac{A_0}{\sqrt{2 \pi } \sigma }\,e^{-\frac{t^2}{2 \sigma ^2}} \; e^{i \omega_0 t},
\label{eqa:GaussFunct}
\end{equation}
where $A_0$ is the amplitude of the field and $2.355 \, \sigma$ is the full width at half maximum (FWHM) of the Gaussian pulse. As is well known, the Fourier transform of a Gaussian function is still a Gaussian with a FWHM inversely proportional to that of the original one. Hence, the envelope of the Gaussian excitation in the frequency domain reads:

\begin{equation}
\begin{array}{r@{\ }c@{\ }l}
E_{in0}[\omega]&=&\frac{1}{\sqrt{2\pi}} \int_{-\infty}^{\infty}E_{in0}[t]\,e^{-i\,\omega\,t}\\
& \vdots & \\
&=&\frac{A_0}{\sqrt{2 \pi }} \; e^{-\frac{1}{2} \sigma ^2 \omega ^2}.\\
\end{array}
\label{eqa:Gauss1Funct}
\end{equation}
Therefore, by increasing the pulse duration the spectral width is decreased.

Interestingly, the general solution presented in subsection \ref{sec:generalResolution} using the Green's function leads straightforwardly to the response of the system in the case of the Gaussian excitation:
\small
\begin{equation}
\begin{array}{r@{\ }c@{\ }l}
E_{out}[t]& = &  A_0 \; e^{i \omega_0 t} \left\{\frac{e^{-\frac{t^2}{2 \sigma ^2}}}{\sqrt{2 \pi } \sigma } - e^{\frac{1}{2} (\gamma +\Gamma ) \left(\sigma ^2 (\gamma +\Gamma )-2 t\right)}  \right.\\
& & \left. \times \Gamma \left(1 + \mbox{erf} \left[ \frac{t - \sigma^2 (\gamma + \Gamma)}{\sqrt{2} \sigma} \right] \right)\right\}. \\
\end{array}
\label{eqa:EoutGaussian}
\end{equation}
\normalsize
where $\mbox{erf}[z] = \frac{2}{\sqrt{\pi}} \int_0^z e^{-t^2} dt$ is the \emph{error function}\cite{SpecialFunctionFor}.

\subsection{FDTD simulation}

In order to illustrate the results given by the previous analytical model, we perform a FDTD simulation \cite{taflove_computational_2010, taflove_advances_2013}, using Meep, an open-source software package \cite{oskooi_meep:_2010}.
As in the first section, we study a straight waveguide coupled to a microring resonator.
A preliminary step consists in computing a resonant frequency of the system. To do so, the microring is excited with a Gaussian source with a relatively large frequency span. Then, the function \emph{harminv} determines the quality factors of different modes. Finally, to study the time-response of the resonator to a pulse of light, a continuous wave source with frequency $f_ {cen}$ corresponding to the best quality factor is inserted at the bus input at $t = 0$ and stopped at $t = 3000\,a.u$.

The different cases of the analytical study are reproduced by keeping the geometry constant and varying the imaginary part of the refractive index of the microring. Doing so, the extrinsic quality factor remains constant, whereas the intrinsic one is tuned to achieve the different coupling regimes.

% RESULTS SECTION ========================================================================
\section{Result discussion}
\subsection{Output intensity in the frequency domain}  
\label{sub:frequency}
%------------------------------

Let us analyze the different system responses in the usual frequency domain. The output electric field intensity as a function of frequency assumes the conventional symmetric Lorentzian shape given by equation \ref{eq:tr} and, depending on the values of the extrinsic $\Gamma$ to intrinsic $\gamma$ losses, three coupling regimes can be identified: critical-($\Gamma=\gamma$), under-($\Gamma<\gamma$) and over-coupling ($\Gamma>\gamma$). 

\begin{figure}[!t] %%ht!
	\begin{center}
		\centering\includegraphics[width=8.3cm]{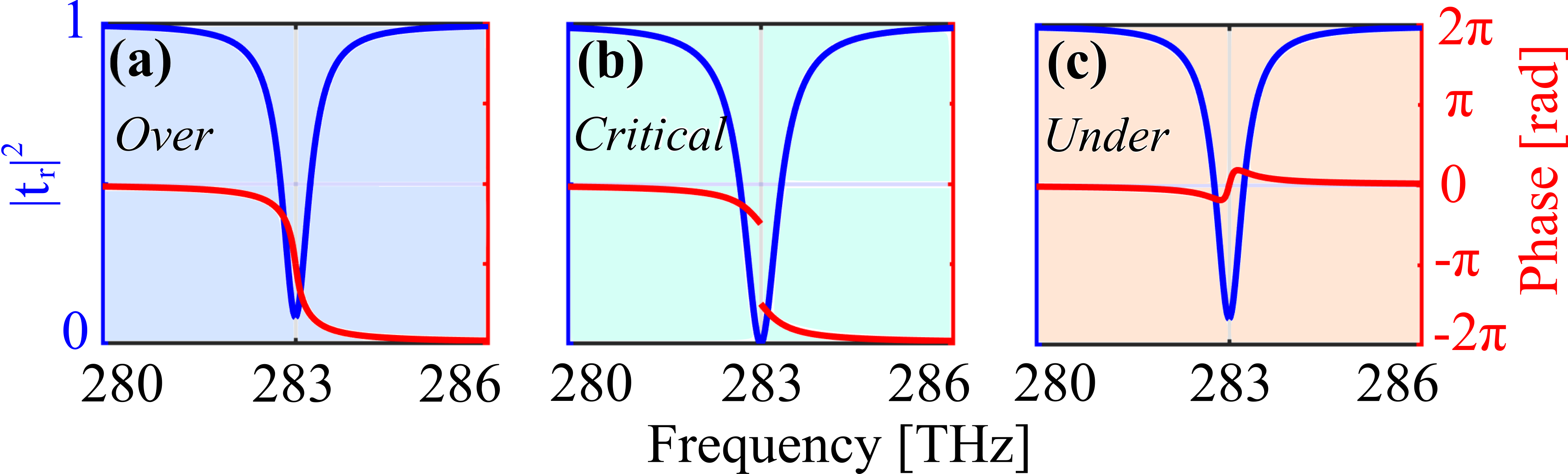}
		\caption{Intensity and phase of the output transmission response ($t_r$) as a function of the frequency. Panels (a), (b) and (c) display the over-, critical- and under-coupling regime. The parameters used are $\omega_0=283\,THz$, $\Gamma\,(\gamma)=162\,GHz$, $\gamma\,(\Gamma)=90\,GHz$ for the over(under)-coupling regime and $\Gamma=\gamma=162\,GHz$ for the case of critical-coupling.  }
		\label{fig:OutputField}
	\end{center}
\end{figure}

Figure \ref{fig:OutputField} (a), (b) and (c) displays the intensity ($|t_r|^2$) and phase ($arg[t_r]$) of the output electric field as a function of the frequency for the three regimes: over-, critical- and under-coupling respectively. The Lorentzian of the under-coupling regime of figure \ref{fig:OutputField} (c) is determined by swapping the value of the extrinsic and intrinsic coefficient of the over-coupling case of panel (a). It is worth noticing that the two Lorentzians are identical. Thus it is not possible to  discriminate in (a) and (c) these two regimes by only knowing the transmission in the frequency domain.  

Equation \ref{eq:Eout} shows that the frequency transmission response is determined by the interferences between the field coming out of the resonator and the field which goes through the bus waveguide. Due to the resonance condition, on resonance, these two fields acquire a phase difference equal to $\pi$. As will be explained in detail in the next section, this difference in phase is crucial to understand the coupling regime of the system. In particular, as shown in figure \ref{fig:OutputField}, the phase shift on resonance reduces to $\pi$ ($0$) in the case of over(under)-coupling regime.

\subsection{Output intensity in the time domain} % -------------------------------

\begin{figure}[!t]
	\centering
	\includegraphics[width=8.2cm]{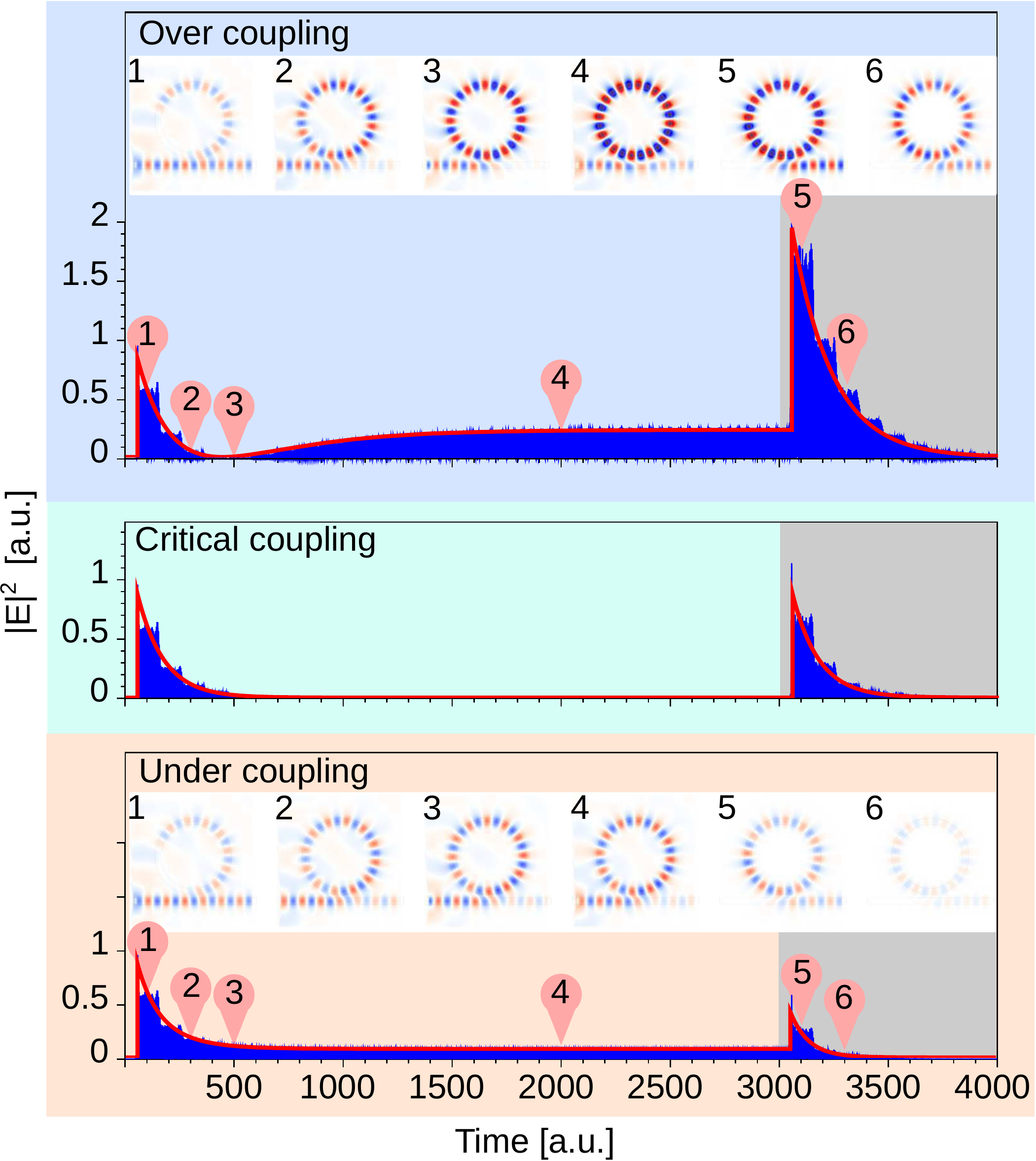} 
	\caption{FDTD simulations of a microring coupled to a bus waveguide excited with a resonant CW source. The source is turned on at $t= 0$ and stopped at $t = 3000\, a.u$. which is represented by the gray background. The gap remains constant while the intrinsic losses are varied to reproduce the three coupling regimes. The plots show the squared electric field norm at the output of the waveguide during the time computed with the analytical model (red) and from the FDTD simulation (blue). The images show the electric field $z$-component, $E_z$. The same arbitrary units are used in the three regimes to allow comparison.}
	\label{fig:MeepResults}
\end{figure}

Let us discuss the output electric field as a function of time for the three coupling regimes over-, critical and under-coupling, considering the analytical model and the FDTD simulation described in the previous section. The three plots in figure \ref{fig:MeepResults} show the squared norm of the electric field at the bus output as a function of time for the three coupling regimes. For the first and last ones, we also report images, obtained from the FDTD simulation, of the electric field z-component distribution at the different times indicated by the red pins in the squared field temporal profiles. The same arbitrary units are used which allow a direct comparison of the three regimes.

Let us describe the different sequences in the \emph{over-coupling} case (first panel of figure \ref{fig:MeepResults}). 

\textbf{Time 1}: After a small delay needed for the light to propagate along the bus waveguide, the power is detected at the output. Its value is the source's one minus the part that is coupled into the ring. The image number 1 shows the field leaving from the bus and the one that begins a round trip inside the microring after being injected by evanescent coupling. The field in the second part of the bus is slightly weaker than the one in the first part. 

\textbf{Time 2}: After three round-trips, the exiting power results from the interferences between the field propagating in the bus and the one exiting from the ring. Image number 2 shows that these interferences destroy the field, hence the exiting power decreases after each round-trip as the field gets stronger and stronger in the resonator. As expected, the magnitude of the output electric field decreases for increasing the time, following an exponential shape until the system reaches the time 3. This exponential behavior is given by the charge of the resonator.

\textbf{Time 3}: After four round-trips, the field trapped in the resonator is large enough to make the interferences able to completely destroy the exiting wave.

\textbf{Time 4}: After a few more round-trips, the field inside the resonator becomes too large and the interferences are not balanced anymore. The wave exiting from the resonator prevails and a power is detected again at the output of the bus. We can also observe that the stationary regime is reached (corresponding to the minimum of the spectral Lorentzian of figure \ref{fig:OutputField} (a)), which means that the power losses in the resonator are exactly balanced by the coupled power.

\textbf{Time 5}: Once the source has been turned off, the interferences can not take place anymore and the power detected at the output is directly the one coming from the resonator. Since the power trapped in the resonator is larger than the source's one, the power detected at the output is larger than the one detected at time 1.

\textbf{Time 6}: At each round-trip, the microring discharges the trapped power, thus, the exiting power fades away as time goes by. Then the magnitude of the output electric field decreases by the increasing of time, following the same exponential law of the charge case. This description of the time evolution of the electric field in the particular case of the over-coupling regime can also be found in \cite{heebner_optical_2008}.

In the \emph{critical-coupling} case, the stationary regime corresponds to the situation when the interferences exactly destroy the exiting wave. Therefore, after a few moments needed to charge the microresonator, no power is detected at the output anymore. 

It is worth noting that, when the output power vanishes, it means that the amplitudes of the electric fields
going through the bus and exiting from the resonator are equal. Meanwhile, it does not imply that
all of the power coming from the source is coupled into the resonator.

On the other hand, in the \emph{under-coupling} regime, the round-trip losses become equal to the coupled power before the power inside the resonator becomes large enough to observe the destroying interferences. Thus, the stationary regime is reached without observing the canceling of the exiting power.
At time 5, when the source is turned-off, the exiting power is smaller than the one measured at time 1 because  the power trapped in the microring is smaller than the source's one.

As expected, also in the critical- and under-coupling regime the charge and discharge of the resonator are characterized by a typical exponential trend.  Interesting is the fact that just measuring the intensity of a resonance response of the system as a function of the time, one can determine the coupling regime. As shown in the subsection \ref{sub:frequency}, this is not possible in the frequency domain.

Our simple analytical model can perfectly predict the simulation behavior observed. The red lines of figure \ref{fig:MeepResults} show the analytical fit of the numerical simulation by using equation \ref{eqa:Eout}. As the extrinsic coefficient is connected to the coupling between the bus waveguide and the resonator, it is equal to $\Gamma=20800\,a.u.$ in all three cases. Instead, the intrinsic coefficient is about $\gamma=6400\,a.u.,\,\gamma=20800\,a.u.$ and $\gamma=38800\,a.u.$ for over-, critical- and under-coupling regimes respectively.  

% we fix the scale of times as 3000 a.u. == 30 ps. In this way we obtain GHz for the intrinsic and extrinsic coefficient, with a radio of the ring of about 10^2 um. Then we multiply for 10^2 in order to write the fit coefficients in a.u. 

%------------------------------------------------

\subsection{Phase shift}

The dynamics of the charge and discharge of the resonator can be understood by looking at the output electric field. The phase is strictly connected to the interference between the two fields in the coupling region: one which exits from the resonator and one which propagates in the bus waveguide. As a result, the phase contains information about the coupling regime of the system. In particular, on resonance and in the stationary regime, the accumulated phase is $0$, $\pi$ in under- and over-coupling regime, respectively.

This can also be deduced from the FDTD simulations.
In figure \ref{fig:MeepResults}, if we concentrate on the field image 4 we can note that the relative phases of the fields are different in the two regimes. Consider the sign of the field $E_z$ given by the colors (blue negative, red positive). In the coupling region and in both cases, $E_z$ propagating in the bus waveguide is positive while $E_z$ exiting from the ring has suffered a $\pi$ phase shift and therefore is negative. Since in the over-coupling regime, the field from the ring is dominating the interference, the field transmitted through the coupling region acquires a $\pi$ phase shift. On the contrary, in the under-coupling regime, the field in the bus waveguide is dominating the interference and, therefore, the transmitted field has no phase shift. 

\begin{figure}[!t] %%ht!
	\begin{center}
		\centering\includegraphics[width=8.2cm]{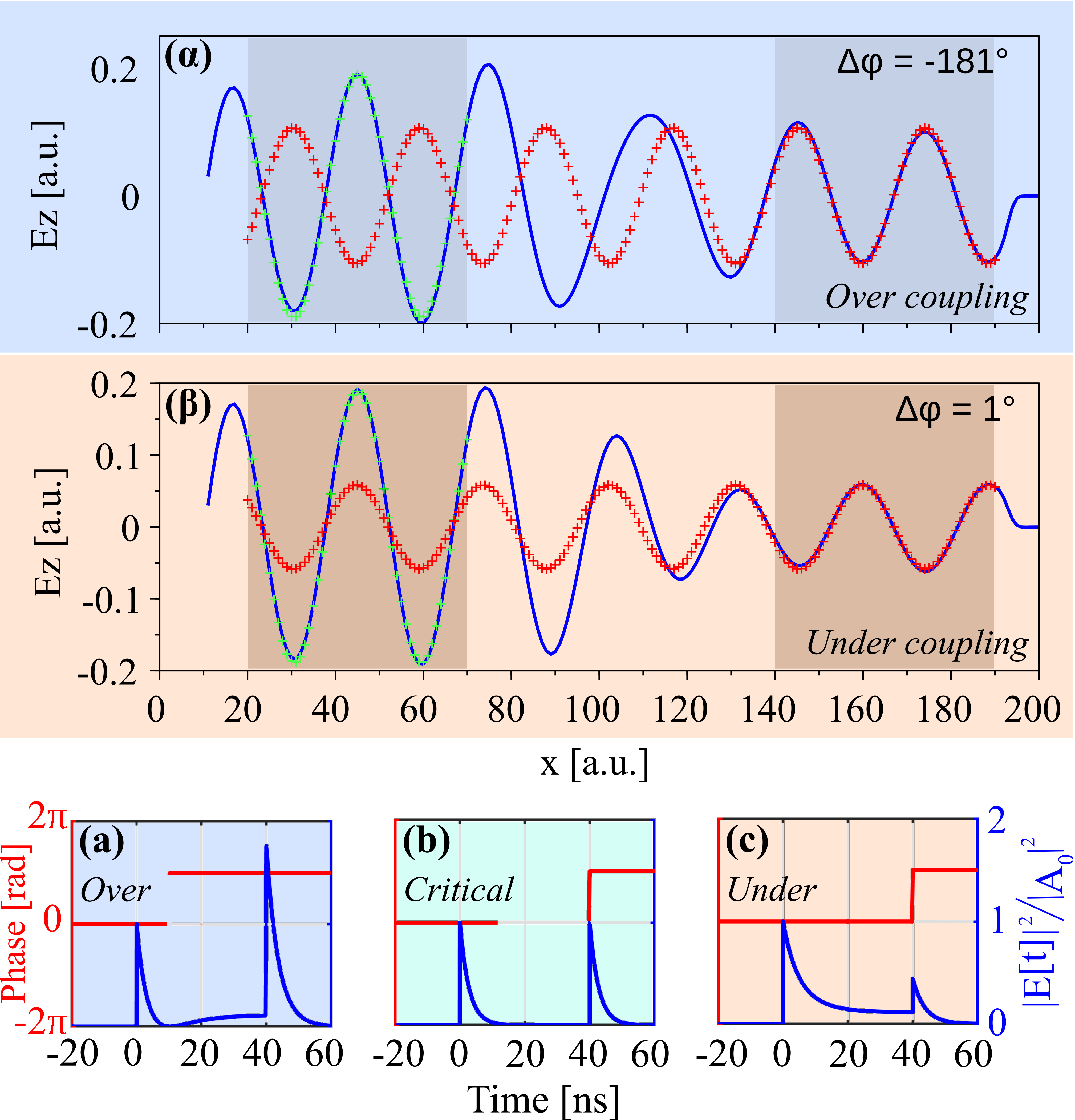}
		\caption{Electric field $z$-component along a cut-line in the middle of the bus waveguide simulated by FDTD at $t = 2000\,a.u$. corresponding to the stationary regime in the over- ($\alpha$) and under-coupling ($\beta$) regimes. An harmonic fit at the input ($(20 < x < 70) \,a.u.$, green crosses in the dark area) and the output ($(140 < x < 190) \,a.u.$, red crosses in the dark area) allows to evaluate the phase shift $\Delta \varphi$ of the output wave in respect to the input. Panels (a), (b) and (c) display the analytical intensity and phase of the output electric field norm as a function of the time in the different coupling regimes.}
		\label{fig:OutputFieldPhase}
	\end{center}
\end{figure}

A more accurate illustration of the phase shift is given for the stationary regime in figure \ref{fig:OutputFieldPhase}.
($\alpha$) and ($\beta$) show the electric field z-component along a cut-line in the middle of the bus waveguide at $t = 2000\,a.u$. of the simulation (i.e. time 4 in figure \ref{fig:MeepResults}) in the over- and under-coupling regime respectively.
To determine the phase of the output wave with respect to the input, a sinusoidal fit was made on the left-hand side of the waveguide ($(20 < x < 70)\,a.u.$, green crosses in the gray area) and on the right-hand side ($(140 < x < 190)\,a.u.$, red crosses in the gray area). We can see that in the over-coupling regime the output wave has a $\pi$-phase shift whereas it is not the case in the under-coupling case. The $1^o$ difference between the theoretical and simulated phase shift is due to the size of the simulation cell. Its small extension forbids to select the fit interval further away from the resonator.

Interestingly, the  analytical model is also able to correctly display the phase behavior. Figure \ref{fig:OutputFieldPhase}(a)-(c) displays the typical intensity of the electric field and the phase as a function of time for the different regimes: over-, critical- and under-coupling (panels (a), (b) and (c) respectively). During the resonator discharge part, all the cases exhibit a phase shift of $\pi$ (last exponential part of figure \ref{fig:OutputFieldPhase}(a)-(c)). This is explained by the fact that only the field exiting from the resonator survives when the input field is turned off. This field has accumulated, due to the resonance condition, a $\pi$ shift with respect to the input electric field. In the under-coupling regime, the exciting (input) field is always dominant over the field which exits from the resonator so that the phase shift is $0$ during the charge and in the stationary condition (see figure \ref{fig:OutputFieldPhase}(c)). In the critical-coupling, the phase is $0$ until the field reaches the stationary condition where the amplitude of the field coming out from the resonator is equal to the exciting one and, therefore, no field is transmitted due to the complete destructive interference (see figure \ref{fig:OutputFieldPhase}(b)). More interesting is the over-coupling regime. As show in figure \ref{fig:OutputFieldPhase}(a), during the resonator charging, the phase of the output field is 0. At some time, the system reaches the perfect destructive interference condition (see image 3 in figure \ref{fig:MeepResults}). Thereafter, the field coming out from the resonator is dominant over the input one so that the accumulated phase is $\pi$ and the system reaches the stationary condition.

%---------------------------------------------------------------------------------
\subsection{Q-factor determination}
In all cases of figure \ref{fig:MeepResults}, the exponential trend at the input field switch off/on is connected to the quality factor of the system. The frequency-dependent definition of the quality factor $Q$ is \cite{heebner_optical_2008,Jackson}:

\begin{equation}
Q[\omega]=\omega\,\frac{\xi}{P_{loss}},
\label{eqa:QualityFactor}
\end{equation}
where $\xi$ is the energy stored inside the resonator, $P_{loss}$ is the power loss, $\omega$ is the angular frequency at which the stored energy and the power loss are measured. From this definition one can obtain the following equation:

\begin{equation}
\xi[t]=\xi[0]\,e^{-\frac{\omega}{Q}\,t},
\label{eqa:EnergyDecay}
\end{equation}
where $\xi[0]$ is the energy stored in the resonator at $t=0$. This exponential decay can be related to our analytical model. In particular, considering the equation \ref{eqa:Eout} one obtains:

\begin{equation}
Q=\frac{\omega_0}{2\,(\Gamma+\gamma)}=\frac{2\,\pi\,f_0}{2\,(\Gamma+\gamma)},
\label{eqa:QinIntAndExt}
\end{equation}
where $f_0$ is the resonant frequency.

%-------------------------------------------------------------------------
\subsection{Interpretation of the steps observed on the simulation}
The FDTD results (blue curve in figure \ref{fig:MeepResults}) exhibit steps which are not observed in the analytical model (red curve). This is due to the fact that the analytical model does not take into account the finite system size. On the other hand, in the FDTD simulation, the output power reaches a new value only when the signal wave coupled to the resonator has performed a complete round-trip which takes about $100\,a.u$.
%In the extreme scenario where the $\Delta t$ is much shorter than the round-trip time ($\tau \ll 100\,a.u.$), the analytical model can not reproduce the transmission spectrum. As shown in fig. [ref], the interference between the field propagating in the bus and the one exiting from the ring can not take place and the response exhibits two distinct peaks.
%\begin{figure}[!t] %%ht!
%	\begin{center}
%		\centering\includegraphics[width=7cm]{dtpiccoloMod.pdf}
%		\caption{a}
%		\label{fig:dtpiccoli}
%	\end{center}
%\end{figure}

\begin{figure}[!t] %%ht!
	\begin{center}
		\centering\includegraphics[width=8.1cm]{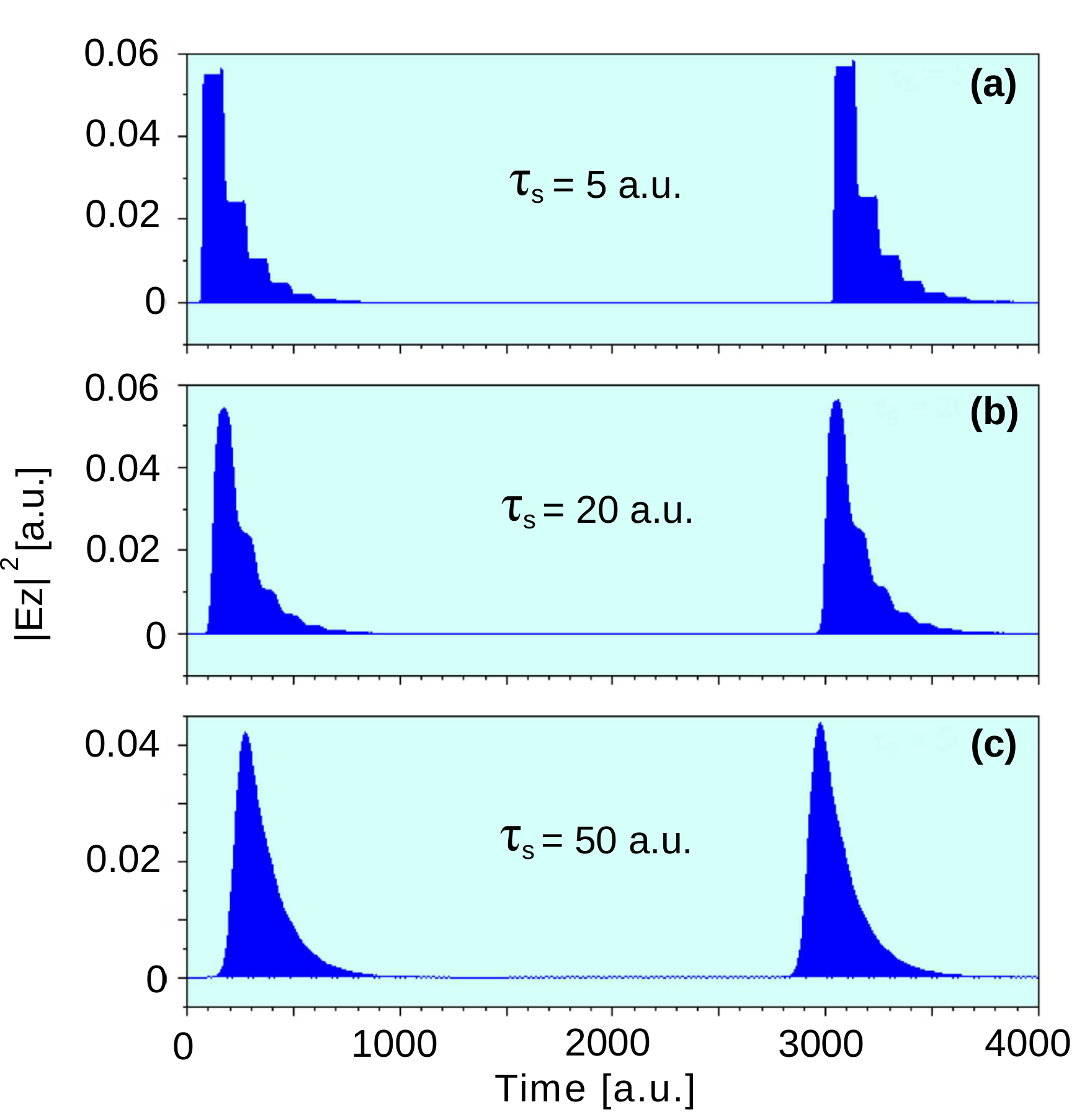}
		\caption{FDTD simulations in the critical coupling case. As in figure \ref{fig:MeepResults}, the source is turned on at $t = 0$ and stopped at $t = 3000\,a.u.$ but, this time, the input pulse has a smooth exponential profile with the time constant $\tau_s$.}
		\label{fig:steps}
	\end{center}
\end{figure}

The sharp response in the FDTD simulation is also related to the temporal profile of the input signal. In the simulation we did, the source was a rectangular pulse, i.e. an abrupt switch-on/off of the signal. This causes the sharp steps. It is possible, instead, to use a trapezoidal pulse with a characteristic on-off exponential edge characterized by a time constant, $\tau_s$. In this case the steps in the simulations are smoothed. Indeed Figure \ref{fig:steps} (a) shows that when $\tau_s$ is small, the steps due to the round-trip delay are clearly visible. As $\tau_s$ increases, the field decrease due to the onset of the destructive interference shows smoother or vanishing steps (see figure \ref{fig:steps} (b) and (c)).

%As $\tau_s$ increases (Fig. \ref{fig:steps} (b-c)), the sudden decreases of the exiting power are attenuated because the front wave has a smooth increasing power. [Or, version 2:] 

%==================================================================================
\section{Experiments}

\subsection{Optical setup}

\begin{figure}[!b] %%ht!
	\begin{center}
		\centering\includegraphics[width=8.1cm]{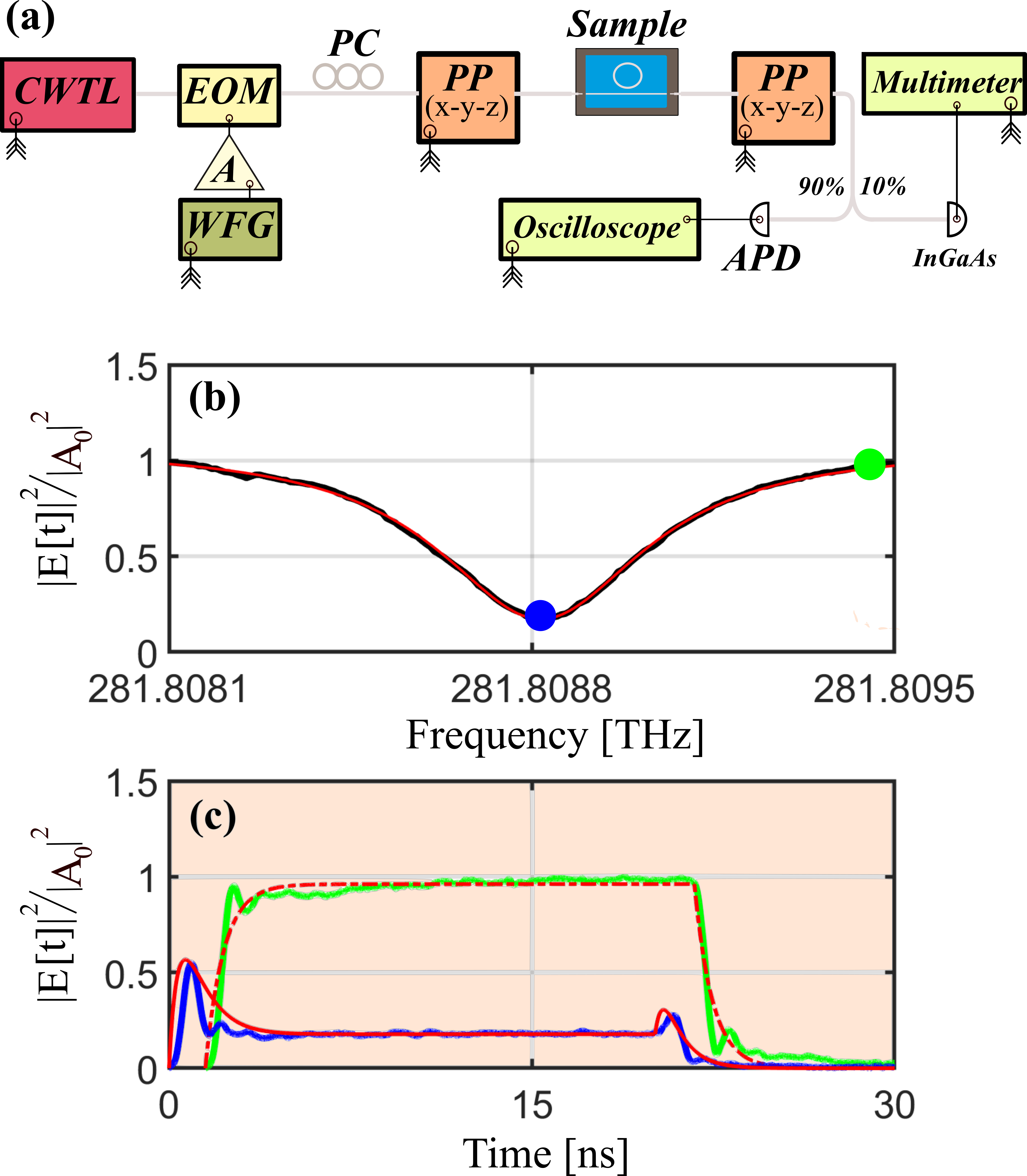}
		\caption{(a) Sketch of the experimental optical setup. CWTL continuous wave tunable laser; EOM electro-optical-modulator; A amplifier; WFG wave front generator; PC polarization controller; PP piezo-positioners; APD silicon avalanche photo diode detector. The symbol ``$< \! \! < \! \! <$'' indicates the remote control via computer. (b) and (c) Experimental norm transmission spectra as a function of the incident frequency and time respectively. The green (blue) data displays the system response out (on) resonance as indicated by the points in panel (b). The red solid-line of panel (a) shows the fit with the Lorentzian of equation \ref{eqa:TrasmissionResponse}. The red dashed-line of panel (b) displays the fit with the typical expression of a $RC$ filter, while the red solid-line shows the plot of the time response using the fit parameter obtained in the frequency spectra. The green data and its fit are shifted in time of about $1.5\,nm$ in order to make the picture clearer.}
		\label{fig:Experimental}
	\end{center}
\end{figure}

Figure \ref{fig:Experimental} (a)  shows our  experimental setup to measure the time response of a microring. The source is a continuum wave tunable laser diode (CWTL) operating in the near infrared range ($1061-1064\,nm$) with a maximum power of about $50\,mW$. An optical fiber guides the beam to an electro-optical-modulator (EOM), which impulses the laser beam. The modulator is controlled via a waveform generator (WFG).
This operates by sending patterns of bits that are sequences of zeros and ones, to a programmable amplifier that drives the optical modulator. Gain and bias of the amplifier are set in order to associate zeros to minimum possible optical output power, and ones to maximum power. The patterns can be as long as $2000\,bits$, while the bit rate can be varied between $50\,MHz$ and $3.2\,GHz$. After the modulator a polarizer controller (PC) controls the polarization of the light. The light pulses are then coupled to the sample via a tapered fiber mounted on a $x-y-z$ piezo-positioners stage (PP). At the end of the sample another fiber mounted on a PP collects the transmitted light signal and transfers it to the detector. This consists of an $InGaAs$ detector or a silicon avalanche photodiode (APD). The $InGaAs$ detector has a much higher sensitivity compared to the APD, however it is much slower and, therefore, it is unsuitable for time resolved measurements. This detector is only used to optimize the coupling between the tapered fibers and the sample, while the APD is used to measure the time dependent response of the system. The APD is used with the internal gain set to about $100$ and no additional electronic amplification.

% * <pierre.guilleme@gmail.com> 2018-11-30T16:38:52.150Z:
% 
% Pulses of light with (?) ns duration and (?) MHz repetition rate are obtained thanks to an electro-optical-modulator controlled via a wavefront generator.
% 
% ^.

\subsection{Results}
% Q in frequency different with respect to Q in time
The samples studied are $Si_3N_4$ ultra-high-quality factor ring resonators monolithically integrated on a silicon chip. In particular, we have measured the response of strip-loaded ring resonator with a radius of $350\,\mu m$. More details on the samples are reported in \cite{Stefan:15}. 

Figure \ref{fig:Experimental} (b) displays the transmitted light intensity as a function of the incident frequency. Figure \ref{fig:Experimental} (c) shows the time dependent transmission when a rectangular pulse is used ($20\,ns$ time duration). The green and blue curves represent the data measured when two different laser frequencies are used. As shown in figure \ref{fig:Experimental} (c), the green (blue) line displays the system response out (on) resonance. In order to make the picture clearer, the green data and its fit are time shifted by about $1.5\,ns$. The ring resonator exhibits a time dependent response that is typical of resonators in the under-coupling regime, as we have seen in the first section. 

The experimental data of figure \ref{fig:Experimental} (b) have been fitted using equation \ref{eqa:TrasmissionResponse}. As one can see in figure \ref{fig:Experimental}  (c), the field far from the resonance frequency (green line) is not perfectly steep as in the theoretical model of section \ref{Sec:Theory}. This behavior is related to the response times of the EOM and APD detector, which introduce two additional time constants, with the EOM one being negligible ($10\,GHz$ BW). As a result, the system exhibits a smooth response. The response time of the experimental set-up can be estimated by fitting the signal out of resonance with the convolution between the step function of equation \ref{eqa:StepFunct} and an exponential  ($\frac{\Theta (t) e^{-\frac{t}{\tau }}}{\tau }$).  This convolution assumes the typical expression of a $RC$ filter, where the time constant $\tau$ is connected to the time response of the EOM and of the APD detector \cite{Horowitz:1989}. The same reasoning applies to the transmitted signal on resonance which is given by the convolution between equation \ref{eqa:Eout} and the exponential response ($\frac{\Theta (t) e^{-\frac{t}{\tau }}}{\tau }$). This convolution is plotted in figure \ref{fig:Experimental} (c) by using the parameters obtained from the fit of the signal transmission on resonance ($\Gamma = (5.32 \pm 0.03)\times 10^{8}\,Hz$, $\gamma = (10.0 \pm 0.1)\times 10^{8}\,Hz $, $f_0 = 281.80882\,THz$) and the one of the out of resonance transmitted signal ($\tau = (750 \pm 10)\,ps$). The  model matches the time and the frequency response of the system and allows estimating a quality factor of about $Q\approx (5.74 \pm 0.01)\times 10^5$ at $f_0$ by using equation \ref{eqa:QinIntAndExt}.

%==================================================================================================
\begin{figure}[!t] %%ht!
	\begin{center}
		\centering\includegraphics[width=8.5cm]{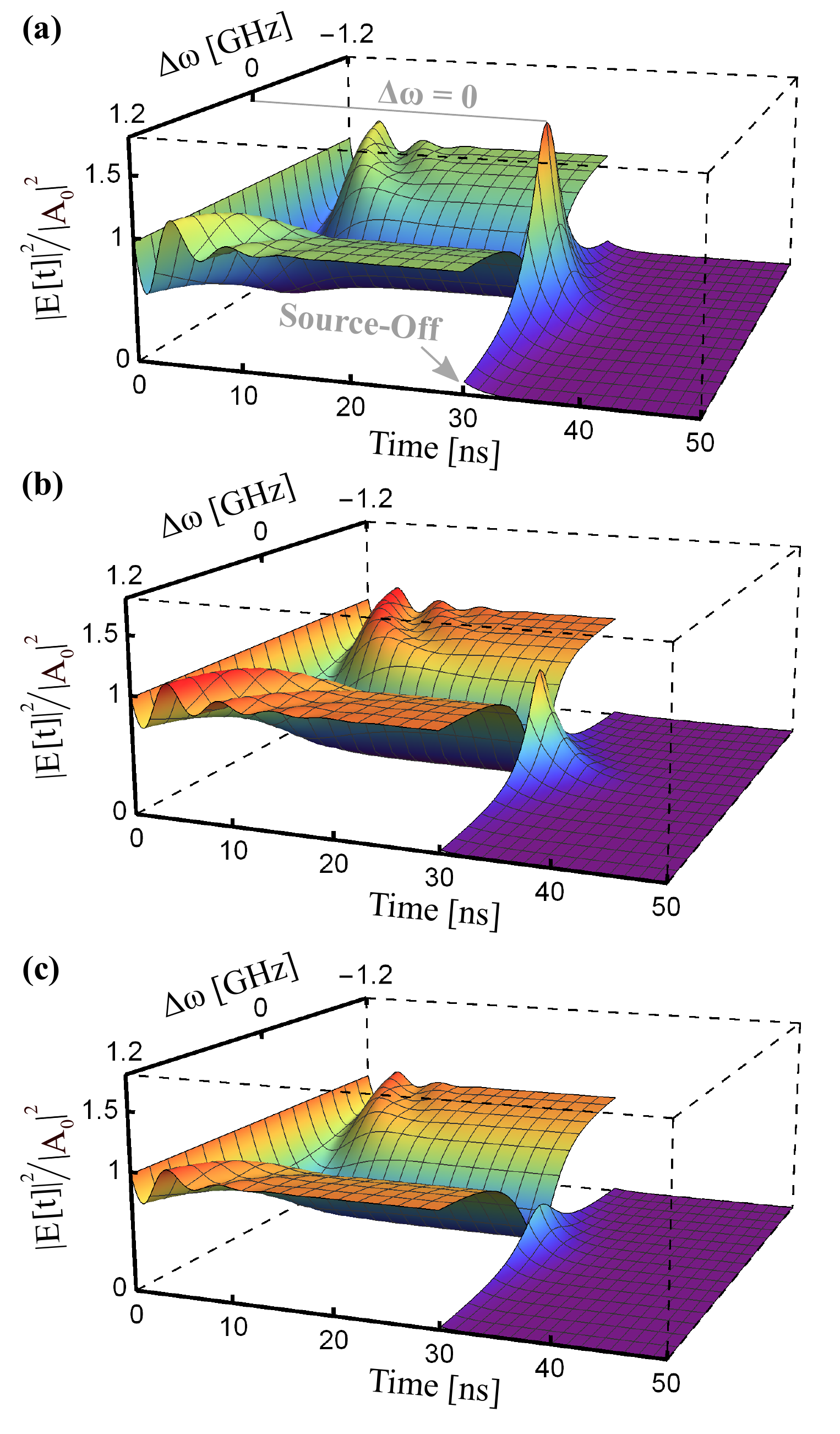}
		\caption{(a)-(c) Norm electric field as a function of the time and frequency. (a) Over-, (b) critical- and (c) under-coupling regime.}
		\label{fig:3dPlot}
	\end{center}
\end{figure} 

\section{Frequency dependence of the time domain output power} \label{sec:freqDependence}

In the previous discussion, we dealt with the particular case in which the frequency of the incident laser is equal to the resonant frequency of the resonator ($\omega_0$). In this section, we take advantage of the general solution given by \ref{eq:EoutGeneral} to discuss the time response of the resonator when the frequency of the incident beam ($\omega$) is varied.

Figure \ref{fig:3dPlot} shows a 3D plot of the output electric field intensity as a function of the time and the frequency detuning ($\Delta \omega$), for the three different regimes: over-, critical- and under-coupling (panel (a), (b) and (c) respectively). Obviously, in these plots, at $\Delta \omega  = 0$, we find again the behavior discussed previously. Since the most efficient energy coupling is achieved for the resonant frequency, the zero-detuning position is easily spotted on the plot by the resonator discharge peak. 

\begin{figure}[!t] %%ht!
	\begin{center}
		\centering\includegraphics[width=8.2cm]{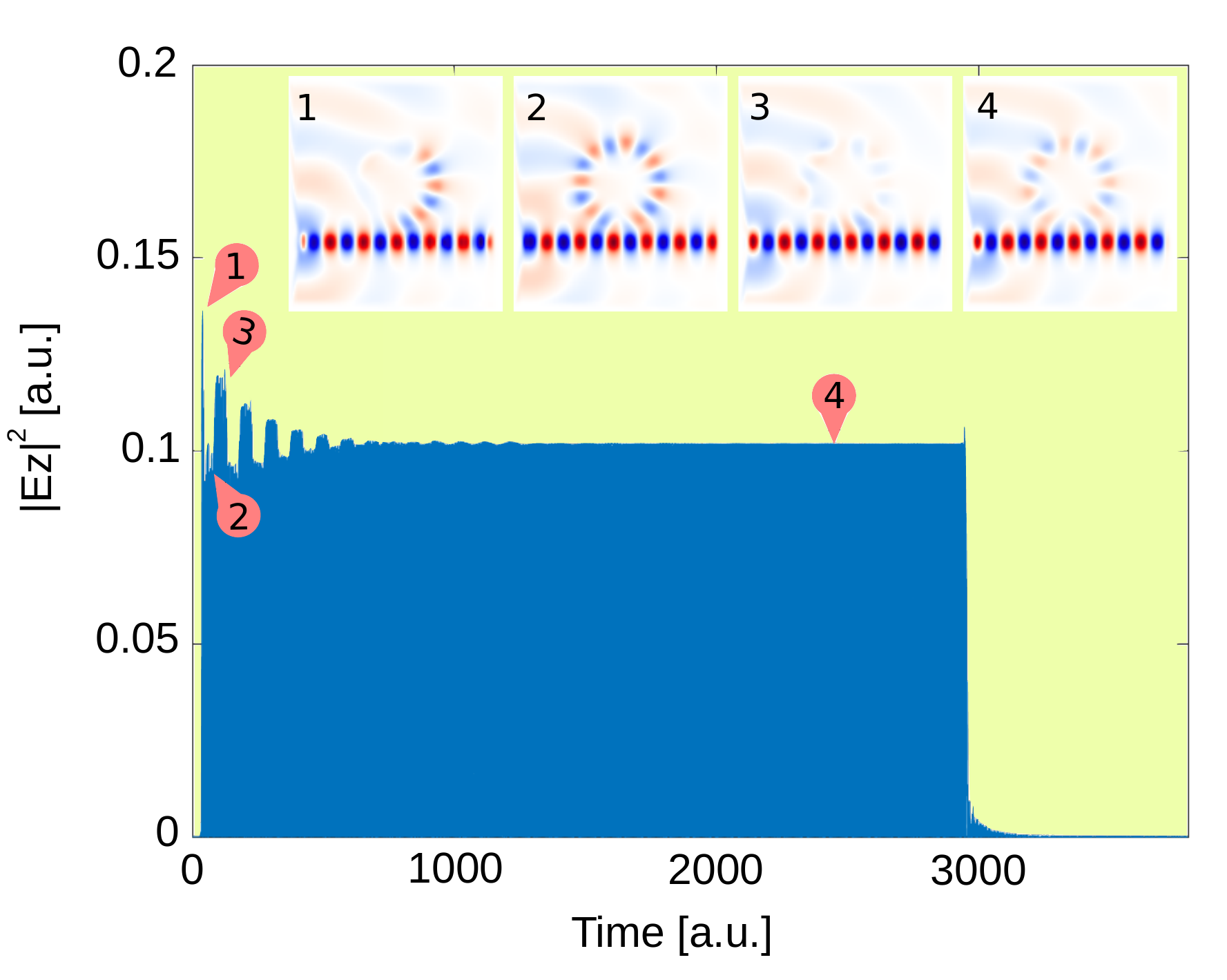}
		\caption{FDTD simulation of a microring coupled to a bus waveguide excited with a CW source of frequency different from the resonant one. The source is switched on at time $t=0\,a.u.$ and off at time $t=3000\,a.u.$. The images show the electric field $z$-component at the time indicated by the red pins.}
		\label{fig:Far}
	\end{center}
\end{figure}

Remarkably, we note in figure \ref{fig:3dPlot} an optical intensity modulation as a function of time, for frequencies far from the resonant frequency. In these cases, the system is still stable but oscillates while approaching the steady-state value. As shown in figure \ref{fig:Far}, this behavior is confirmed by the numerical FDTD simulation, obtained by exciting the system with a rectangular pulse whose frequency is not resonant. Looking at the field patterns we can explain these oscillations.

While in figure \ref{fig:MeepResults}, the constructive interference inside the resonator reinforces the signal field coupled to the resonator at each round trip, here the destructive interference leads to periodic reinforcement and vanishing of the field inside the resonator (images 2,3). In turn, interference in the coupling region between this internal beating wave and the input one induces oscillations in the output field. After several round trips, a stationary regime is reached where few light is coupled inside the cavity (image 4 in figure \ref{fig:Far}).

\section{Conclusion}

In this paper, the response of a resonator excited by a rectangular light pulse has been analyzed. A simple model capable of capturing the physics of the system has been presented. To support the simple analytical model, numerical simulations with an FDTD software have been performed. This study allows to understand the time dynamics of the light propagation in a resonator coupled to a bus waveguide in three different regimes: over- critical- and under-coupling. As a consequence of the coupling regime, the interference between the light propagating in the bus waveguide and that circulating in the microring changes determining both the intensity and the phase delay of the output signal. In particular, it has been shown that the knowledge of the intensity of the time dependent electric field response permits, through its particular evolution, to understand the coupling regime of the system. An experiment demonstrated the validity of the analysis in the case of the under-coupling regime. In this way, the coupling regime and the quality factor of a $Si_3N_4$ ring resonators monolithically integrated on a silicon chip have been estimated.

\section*{Acknowledgement}
I. Carusotto is acknowledged for fruitful discussions and M. Ghulinyan for providing the fine microring resonators. This work was partially supported by ESA under contract 16.1.0293.0.

\ifCLASSOPTIONcaptionsoff
  \newpage
\fi

% biography section
% 
% If you have an EPS/PDF photo (graphicx package needed) extra braces are
% needed around the contents of the optional argument to biography to prevent
% the LaTeX parser from getting confused when it sees the complicated
% \includegraphics command within an optional argument. (You could create
% your own custom macro containing the \includegraphics command to make things
% simpler here.)
%\begin{IEEEbiography}[{\includegraphics[width=1in,height=1.25in,clip,keepaspectratio]{mshell}}]{Michael Shell}
% or if you just want to reserve a space for a photo:

%\begin{IEEEbiography}{Michael Shell}
%Biography text here.
%\end{IEEEbiography}

% if you will not have a photo at all:
%\begin{IEEEbiographynophoto}{John Doe}
%Biography text here.
%\end{IEEEbiographynophoto}

% insert where needed to balance the two columns on the last page with
% biographies
%\newpage

%\begin{IEEEbiographynophoto}{Jane Doe}
%Biography text here.
%\end{IEEEbiographynophoto}

% You can push biographies down or up by placing
% a \vfill before or after them. The appropriate
% use of \vfill depends on what kind of text is
% on the last page and whether or not the columns
% are being equalized.

%\vfill

% Can be used to pull up biographies so that the bottom of the last one
% is flush with the other column.
%\enlargethispage{-5in}

% that's all folks
\end{document}